\documentstyle[12pt]{article}

\newcommand{\resection}[1]{\setcounter{equation}{0}\section{#1}}

\newcommand{\EQ}{\begin{equation}}
\newcommand{\EN}{\end{equation}}
\newcommand{\A}{\begin{array}}
\newcommand{\E}{\end{array}}
\newcommand{\EA}{\begin{eqnarray}}
\newcommand{\EE}{\end{eqnarray}}

\newcommand{\hs}{\hspace{1mm}}
\newcommand{\HS}{\hspace*{1cm}}
\newcommand{\BS}{\hspace*{-1cm}}

\newcommand{\ep}{\varepsilon}
\newcommand{\D}{{\rm d}}

\newcommand{\PR}{Phys. Rev. }
\newcommand{\PRL}{Phys. Rev. Lett. }
\newcommand{\NP}{Nucl. Phys. }

\begin{document}

\topmargin 0pt
\renewcommand{\thefootnote}{\fnsymbol{footnote}}
\newpage
\setcounter{page}{0}

\begin{titlepage}

\begin{flushright}
January 1995
\\ cond-mat/9501094
\end{flushright}

\vspace*{0.8cm}

\begin{center}
{\large \bf ON THE RENORMALIZATION OF THE \\
            KARDAR-PARISI-ZHANG EQUATION} \\

\addtocounter{footnote}{1}
\vspace{1.2cm}
{\large Michael L\"assig}
\\

\vspace{1.2cm}
{\sl Max-Planck-Institut f\"ur Kolloid- und Grenzfl\"achenforschung, \\
     Kantstr. 55, 14513 Teltow, Germany}
\\ {\sl and} \\
{\sl Institut f\"ur Festk\"orperforschung,
     Forschungszentrum, 52425 J\"ulich, Germany}

\end{center}
\vspace{1.2cm}
\setcounter{footnote}{0}

\begin{abstract}
The Kardar-Parisi-Zhang (KPZ) equation of nonlinear stochastic growth
in $d$ dimensions is studied using the mapping onto a system of
directed polymers in a quenched random medium. The polymer problem is
renormalized exactly in a minimally subtracted perturbation expansion
about $d = 2$. For the KPZ roughening transition in dimensions $d > 2$,
this renormalization group yields the dynamic exponent $z^\star = 2$
and the roughness exponent $\chi^\star = 0$, which are exact to all
orders in $\ep \equiv (2 - d)/2$.  The expansion becomes singular in $d =
4$, which is hence identified with the upper critical dimension of the
KPZ equation. The implications of this perturbation theory for the
strong-coupling phase are discussed.  In particular, it is shown that
the correlation functions and the coupling constant defined in minimal
subtraction develop an essential singularity at the strong-coupling
fixed point.
\end{abstract}

\end {titlepage}

\newpage
\setcounter{footnote}{\arabic{footnote}}

\resection{Introduction}

One focus of today's statistical mechanics is scale invariance far from
equilibrium. Driven growth of surfaces is an example that widely occurs in
nature; for a review, see e.g. \cite{KrugSpohn.review}. On large scales of
space and time, the effective growth dynamics may
often be described by a stochastic  evolution equation for a continuous
``height field'' $h(r,t)$. The Kardar-Parisi-Zhang (KPZ) equation \cite{KPZ}
\EQ
\partial_t h = \nu \nabla^2 h + \frac{\lambda}{2} (\nabla h)^2 + \eta
\label{KPZ}
\EN
driven by Gaussian white noise with
\EQ \A{rcl}
\overline{\eta(r,t)} & = & 0 \hs,
\\
\overline{\eta(r,t) \eta(r',t')} & = & \sigma^2 \delta^d (r - r') \delta(t-t')
\E \label{eta}
\EN
has become the ``standard model'' for such processes since it represents the
simplest universality class of {\em nonlinear} growth. Many realistic
growth models are in other universality classes due to additional
symmetries. Moreover, the KPZ equation has deep theoretical links with a
number of more difficult nonequilibrium problems, notably fluid dynamics and
turbulence. Phenomenologically, the link to turbulence is even more manifest
for certain related growth models that show multiscaling
\cite{Krug.turbif}.

The phenomenology of the KPZ equation is well known. In spatial
dimensionalities $d \leq 2$, the nonlinearity $ (\lambda/2) (\nabla h)^2$
is a relevant perturbation of the Gaussian dynamics ($\lambda = 0$). The
strong-coupling regime is characterized by two basic exponents, the {\em
roughness exponent} $\chi$ and the {\em dynamic exponent} $z$, which are
defined e.g. by the asymptotic scaling on large scales
\EQ
\langle (h(r_1, t_1) - h(r_2, t_2))^2 \rangle \sim
|r_1 - r_2|^{2 \chi} {\cal C} (t \,|r_1 - r_2|^{z})
\label{hh}
\EN
of the height difference correlation function \cite{KPZ}.
In the renormalization group, this is a crossover between two fixed points:
the {\em Gaussian fixed point}, which is (infrared--) \linebreak unstable, and
the
{\em strong-coupling fixed point}, which is stable.
In $d = 1$, the exponents $\chi = 1/2$ and $z = 3/2$ can be obtained exactly in
several ways: by exploiting the symmetries of the system (namely Galilei
invariance and  a fluctuation-dissipation relation particular to $d = 1$), by a
one-loop dynamic renormalization group analysis \cite{FNS.Burgers}, or by
mapping the KPZ dynamics onto an exactly solvable lattice model
\cite{GwaSpohn.Bethe}. All of these tools fail in higher
dimensions, and the properties of the strong-coupling fixed point are known
only numerically.
In $d = 2$, the KPZ equation is asymptotically free, and the  crossover to the
strong-coupling regime is exponentially slow
\cite{NattermannTang.weakcoupling}. Recent numerical values for the exponents
are $\chi = 0.386$ and $z = 1.612$ \cite{Ala-NissilaAl.num}.
For $d > 2$, the height profile is smooth in the Gaussian theory. A small
nonlinearity $ (\lambda/2) (\nabla h)^2$ does not alter this asymptotic
scaling; there is now a {\em roughening transition} to the
strong-coupling phase at finite critical values $\pm \lambda_c$
\cite{ImbrieSpencer,CookDerrida,EvansDerrida}. In the
renormalization group, the transition is represented by a third fixed point.
This {\em critical fixed point} is unstable and appears between the Gaussian
fixed point and the strong-coupling fixed point which are now both stable
\cite{NattermannTang.weakcoupling}. Numerical studies
\cite{num,Ala-NissilaAl.num} indicate that a strong-coupling
phase with $z < 2$ persists also in high dimensions; various theoretical
arguments, on the other hand, predict the existence of a finite upper critical
dimension $d_>$, above which $z = 2$ in both the weak and strong coupling
regimes
\cite{CookDerrida,HalpinHealy,FeigelmanAl}.

A satisfactory theory of stochastic growth should classify the different
universality classes and the possible crossover phenomena between them, as well
as give a way to calculate scaling indices exactly or in a controlled
approximation. Despite considerable efforts, such a theory still seems far.
In the framework of the renormalization group, the strong-coupling fixed point
does not seem to be accessible by the methods of renormalized perturbation
theory and the $\ep$-expansion that have been so successful in equilibrium
critical phenomena. This key difficulty is a further common feature of the KPZ
equation and turbulence, and one may speculate that its eventual solution will
be similar in both cases as well.

Recent progress has taken place mainly along a different avenue. The numerical
solution of the so-called mode-coupling equations, a self-consistent
approximation to the full problem,  produces exponents which  are not very far
from the best numerical estimates for the KPZ equation
\cite{BouchaudCates,DohertyAl,Tu,MooreAl}. Moreover, the mode-coupling
equations have
been shown to be exact in a certain  large-$N$ limit \cite{DohertyAl}. However,
for two reasons it is not clear at present whether the mode-coupling  solution
can serve as the basis for a controlled expansion: (a) The numerical solution
of the mode-coupling equations presents great difficulties and e.g.~for $d =
2$,
it relies so far on assumptions on the approximate scaling form of the
propagator. (b) The transition to finite $N$ may well introduce new
singularities that have to be handled by some kind of renormalization.

The aim of this paper is to compare the strong-coupling phase with the
roughening transition under renormalization group aspects. The critical
fixed point is perfectly accessible in an $\ep$-expansion about the
lower critical dimensionality $ d_< = 2$, as I show in sections 2 and 3
by dynamic renormalization and by exploiting the mapping of the KPZ
equation onto a system of directed polymers with quenched disorder. The
structure of the perturbative singularities is in fact very simple. In
terms of the dimensionless coupling constant $u_M^2$ in a minimal
subtraction scheme, the beta function reads exactly to all orders in
perturbation theory
\EQ
\beta (u_M^2) = 2 \ep u_M^2 + 2 (u_M^2)^2
\label{beta}
\EN
with $\ep = (2 - d) / 2$. This yields the dimension-independent critical
exponents
\EQ
z^\star = 2 \hs, \HS \chi^\star = 0
\label{chistar.zstar}
\EN
at the roughening transition, which agree with a scaling argument by Doty
and Kosterlitz \cite{DotyKosterlitz} and with a one-loop dynamic
renormalization group calculation \cite{NattermannTang.weakcoupling} that
has recently been extended to two-loop order \cite{FreyTauber}.

The perturbation theory for the roughening transition is likely to be
exact in the interval $ 2 \leq d \leq 4 $; however, $d = 4$ is seen to be
a singular point. It is hence tempting to identify $d_> = 4$; see the
discussion
below.

In sect.~4, I turn to the consequences of this $\ep$-expansion for the
crossover to the strong-coupling fixed point in $d = 2$. It proves necessary
to carefully distinguish between fields and couplings in {\em minimal
subtraction} and properly {\em renormalized\/} fields and couplings defined by
their finiteness at a renormalization point. In particular, it is shown that
the functional dependence of the renormalized coupling constant $u_R$ and the
renormalized height field $h_R$ on their minimal subtraction counterparts,
\EQ
u_R (u_M) \hs, \HS
h_R (h_M, u_M) \hs,
\EN
has an essential singularity at $u_M = 0$. (In an ordinary $\ep$-expansion, the
mapping $u_R (u_M)$ is a diffeomorphism of which both the ultraviolet fixed
point $u_R = u_M = 0$ and the infrared fixed point $u_R^\ast (u_M^\ast)$ are
regular points.) This property allows to pin down the reasons for the failure
of perturbation theory for the strong-coupling fixed point.

The results are summarized and discussed in sect.~5.

\resection{The Roughening Transition: Dynamic \newline Renormalization}

In this section, I will sketch the dynamic renormalization of the KPZ
equation
\cite{FNS.Burgers,NattermannTang.weakcoupling,SunPlischke.kpz,FreyTauber}
in a formalism that facilitates comparison with the renormalization for the
polymer
system.

It is convenient to use the dynamic functional \cite{dynamic.functional}
\EQ
\int {\cal D} h {\cal D} \tilde h
     \exp \left [ - \int \D^d r \D t_0 \left (
    \frac{1}{2} \tilde h_0^2 + i \tilde h_0 \left (
    \frac{\partial}{\partial t_0} h_0 - \frac{1}{2} \nabla^2 h_0
    - \frac{\lambda_0}{2} (\nabla h_0)^2  -\rho  \right ) \right )
          \right ]
\label{dynaction}
\EN
in terms of the field $h_0$ and the ``ghost'' field $\tilde h_0$, which
generates response functions,
\EQ
\left \langle \prod_{j = 1}^{\tilde N} i \tilde h_0 (r_j, t_{0j})
              \prod_{j = \tilde N + 1}^{\tilde N + N} h_0 (r_j, t_{0j})
\right \rangle =
\prod_{j=1}^{\tilde N} \frac{\delta}{ \delta \rho (r_j, t_{0j})}
\left \langle \prod_{j = \tilde N + 1}^{\tilde N + N} h_0(r_j, t_{0j})
\right \rangle \hs.
\EN
Here the convention has been adopted to absorb all dimensionful constants of
the linear theory into the ``canonical'' variables
\EQ
t_0 = \nu t \hs, \HS
h_0 = \left ( \frac{\nu}{\sigma^2} \right ) ^{\frac{1}{2}} h \hs, \HS
\tilde h \hs,
\EN
which have dimensions
\EQ
z_0 = 2 \hs, \HS
- \chi_0 = \frac{d - 2}{2}  \hs, \HS
\chi_0 + d = \frac{d + 2}{2} \hs,
\EN
respectively. This convention is standard in field theory, but unfortunately is
not generally  used in the literature on dynamic renormalization, which tends
to burden the  calculations with redundant factors.

The linear theory has the response propagator
\EA
\lefteqn{ \BS \BS G_0 (r_2 - r_1, t_{02} - t_{01}) \equiv
\langle i \tilde h_0 (r_1, t_{01}) h_0 (r_2, t_{02}) \rangle =  } \nonumber
\\ & & \HS
\frac{ \theta (t_{02} - t_{01}) }{ (4 \pi (t_{02} - t_{01}))^{-d/z_0} }
\exp \left [ \frac{ - (r_2 - r_1)^{z_0} }{ t_{02} - t_{01} } \right ]      \hs.
\label{G0}
\EE
The formal expression for the height-height correlation function
\EA
\lefteqn{ C_0 (r_2 - r_1, t_{02} - t_{01}) \equiv
\langle h_0 (r_1, t_{01}) h_0 (r_2, t_{02}) \rangle =  } \nonumber
\\ & &
\int \D^d r \D t_0 \,
     G_0 (t_{01} - t_0, r_{01} - r_0) G_0 (t_{02} - t_0, r_{02} - r_0)
\label{C0}
\EE
requires for $d < 2$ (i.e. $\chi_0 > 0$) the introduction of an infrared
cutoff. In a system of finite size $L$ with periodic boundary
conditions, the stationary correlation function at late times $t_1, t_2$
is  translationally invariant; it has a singularity
\EQ
C_0 (r_1 - r_2, t_{01} - t_{02}, L) \sim L^{2 \chi_0} \hs.
\EN
Only the infrared-regularized correlation function remains well-defined
in the thermodynamic limit $L \to \infty$:
\EA
\lefteqn{\BS \; C_0' (r_2 - r_1, t_{02} - t_{01}) \equiv
         \lim_{L \to \infty} C_0 (r_2 - r_1, t_{02} - t_{01}, L) - C_0 (0,0,L)
=   }
\nonumber \\ & & \HS \HS \HS \HS
|r_2 - r_1|^{2 \chi_0}  F ( |t_{02} - t_{01}| / |r_2 - r_1|^{z_0} ) \hs.
\label{C0'}
\EE
In the interacting theory, the same subtraction is necessary to define
the $L$-indepen-dent stationary two-point function $C'(r, t_0, u_0)$;
the higher connected correlation functions require infrared
regularizations as well. The scale $L$ will also serve to generate the
renormalization group flow below.

The canonical coupling constant
\EQ
\lambda_0 = \left ( \frac{\sigma^2}{\nu^3} \right )^{\frac{1}{2}} \lambda
\EN
has the dimension $\ep \equiv (d - 2) / 2$. We define the dimensionless
coupling constant
\EQ
u_0 = \lambda_0 L^{\ep}  \hs.
\EN
The response and correlation functions of the nonlinear theory have the
crossover scaling form
\EQ
\left \langle \prod_{j = 1}^{\tilde N} i \tilde h_0 (r_j, t_{0j})
          \prod_{j = \tilde N + 1}^{\tilde N + N} h_0 (r_j, t_{0j}) \right
      \rangle =
L^{-N \chi_0 + \tilde N (\chi_0 + d)}
F_{N \tilde N} \left ( \frac{r_j - r_k}{L},
                       \frac{t_{0j} - t_{0k}}{L^{z_0}},
                       u_0
               \right ) \hs,
\EN
which can be expressed as the  ``bare'' Callan-Symanzik equation
\EA
\lefteqn{\BS \BS \BS \BS \;\;
          \left ( L \left. \frac{\partial}{\partial L} \right |_{\lambda_0}
                  + \sum_j r_j \frac{\partial}{\partial r_j}
                  + z_0 \sum_j t_{0j} \frac{\partial}{\partial t_{0j}}
                  + \beta_0 (u_0) \frac{\partial}{\partial u_0}
                  - N \chi_0 + \tilde N (\chi_0 + d)
          \right )  }          \nonumber
\\ & &
\times \left \langle \prod_{j = 1}^{\tilde N} i \tilde h_0 (r_j, t_{0j})
          \prod_{j = \tilde N + 1}^{\tilde N + N} h_0 (r_j, t_{0j}) \right
      \rangle =  0
\label{bareCS}
\EE
with
\EQ
\beta_0 (u_0) \equiv L \partial_L u_0 = \ep u_0 \hs.
\EN
For the infrared-regularized correlators, the explicit dependence on $L$
vanishes in the thermodynamic limit.

In the strong-coupling limit $\lambda_0 \to \infty$, these correlation
functions develop anomalous scaling and hence a singular dependence on
the bare coupling constant $\lambda_0$. Renormalization consists in absorbing
these singularities into new variables
\EQ \A{rcl}
h_R         & = &  Z_h (u_R) h_0  \hs,                  \\
\tilde h_R  & = &  Z_h^{-1} (u_R) \tilde h_0  \hs,      \\
t_R         & = &  Z_t (u_R) t_0  \hs,                  \\
u_R         & = &  Z(u_R) u_0   \hs.                    \E
\label{R}
\EN
Under this change of variables, Eq. (\ref{bareCS}) transforms into the
renormalized Callan-Symanzik equation
\EA
\lefteqn{\BS \BS \BS \BS \;\;\;
          \left ( L \left. \frac{\partial}{\partial L} \right |_{\lambda_0}
                  + \sum_j r_j \frac{\partial}{\partial r_j}
                  + z (u_R) \sum_j t_{Rj} \frac{\partial}{\partial t_{Rj}}
                  + \beta_R (u_R) \frac{\partial}{\partial u_R}
                  - N \chi (u_R) + \tilde N (\chi (u_R) + d)
          \right )  }          \nonumber
\\ & &
\times \left \langle \prod_{j = 1}^{\tilde N} i \tilde h_R (r_j, t_{Rj})
          \prod_{j = \tilde N + 1}^{\tilde N + N} h_R (r_j, t_{Rj}) \right
      \rangle =  0
\label{renCS}
\EE
with
\EA
\beta (u_R) \equiv L \frac{\partial}{\partial L} u_R  & = &
\frac{\ep u_R}{1 - u_R \frac{\D}{\D u_R} Z} \hs,   \label{beta.pert} \\
z(u_R)                           & = &
z_0 - \beta \frac{\D}{\D u_R} \log Z_t \hs,        \label{z}    \\
\chi(u_R)                           & = &
\chi_0 - \beta \frac{\D}{\D u_R} \log Z_h \hs.     \label{zeta}
\EE
A different but equivalent Callan-Symanzik equation is derived in
ref.~\cite{FreyTauber}.
Notice that the renormalization of the ghost field
is not independent since the response function
$ G_R (r_2 - r_1, t_{R2} - t_{R1}, u_R) =
\langle i \tilde  h(r_1, t_{R1}) h(r_2, t_{R2}) \rangle $
always has dimension $d$ by its definition. Furthermore, a Ward identity due to
Galilei invariance \cite{FreyTauber} enforces the following relation between
the $Z$-factors:
\EQ
Z = Z_t^{-1} Z_h^{-1} \hs.
\label{ward}
\EN
By inserting this relation into Eqns. (\ref{beta.pert}), (\ref{z}), and
(\ref{zeta}),
one obtains
\EQ
\beta (u_R) = [ -2 + z (u_R) + \chi (u_R) ] u_R
\EN
and hence at any nontrivial fixed point $u_R^\star \neq 0$ the exponent
identity
\EQ
z (u_R^\star) + \chi (u_R^\star)  = 2 \hs.
\label{scalingrelation}
\EN

The renormalized variables (\ref{R})
can be defined in a nonperturbative
way by imposing two independent normalization conditions e.g. on the
infrared-regularized correlators in an infinite system,
\EA
L^{2 \chi_0} C'_R (L, t_R \! = \! 0, u_R) & = &
L^{2 \chi_0} C'_0 (L, t_0 \! = \! 0) \hs,
\label{CL}
\\
L^d G_R (0, t_R \! = \! L^2, u_R) & = &
L^d G_0 (0, t_0 \! = \! L^2) \hs.
\label{GL}
\EE
$L$ is now an arbitrary normalization scale. An alternative to (\ref{CL}) is
the normalization condition
\EQ
L^{2 - \chi_0} \langle \partial_{t_R} h_R \rangle (L, u_R) = - b \, u_R
\label{hL}
\EN
on the universal finite-size correction to the stationary growth velocity in a
system of
size $L$, where $b > 0$ is a constant independent of $u_R$ that is defined in
(\ref{h1}) below.

In a perturbative $\ep$-expansion, there is an alternative way of constructing
the  $Z$-factors, namely order by order through a minimal subtraction
prescription.  This makes the rhs. of Eqns. (\ref{CL}) and (\ref{GL}) analytic
functions of  the minimally renormalized coupling constant $u_M$ with
coefficients that remain finite as $\ep \to 0$. As long as the coupling
constant is small, this scheme is clearly equivalent to normalization
conditions. As we shall see in sect.~4, this is no longer the case for large
values of $u_R$. I will therefore denote all quantities in the minimal
subtraction scheme by the subscript $M$, and reserve the term ``renormalized''
and the subscript $R$\/ to quantities defined by normalization conditions.

I will discuss the perturbative renormalization not for the
momentum-space response and correlation functions in the infinite
system as it is customarily done but for the position space response
function $G_0 (r, t_0, u_0)$ and the stationary finite-size
amplitude $ \langle \partial_{t_0} h_0 \rangle (L, u_0) $.
 The calculation is in close analogy to the polymer renormalization
 group of the next section.

The response function has the diagrammatic expansion shown in Fig.~1(a). To
order
$u_0^2$, the expansion reads
\EQ
L^d G_0 (0, t_0 \! = \! L^2, u_0) =
L^d G_0 (0, t_0 \! = \! L^2) + O (u_0^2 \ep^0, u_0^4)
 \hs.
\EN
The one-loop diagram does not have a pole at $d = 2$ since its short-distance
singularity cancels with a geometric factor $2 - d$
\cite{SunPlischke.kpz,FreyTauber}. As Frey and
T\"auber have shown, constructing the strong-coupling fixed point in $d
= 1$ requires taking into account this ``hidden'' pole of the response
function \cite{FreyTauber}. However, this finite renormalization can be ignored
for
the critical fixed point above $d = 2$, which is the focus of this section.

The expansion for the growth velocity is shown in Fig.~1(b). The tadpole
diagram at order $u_0$ consists of a nonuniversal ultraviolet-divergent
part and of the universal finite-size correction
\EQ
\int_{\;  \rm reg.} \D^d r' \D t'_0 \, [ \partial_{r'} G_0 (r', t'_0, L) ]^2
 = -  b \, L^{-d} \hs,
\label{h1}
\EN
At order $u_0^3$, the boxed subdiagram contributes a pole,
\EQ
L^{2 - \chi_0} \langle \partial_{t_0} h_0 \rangle (L, u_0) =
- b \, u_0 \left (1 +  \frac{c}{\ep} u_0^2 \right ) +
O (u_0^3 \ep^0, u_0^5)
\EN
with $c = 1 / 32 \pi$. This pole originates from the integration region where
the two
vertices approach each other. It can be absorbed into the definition of the
variables
$h_M = Z_{Mh} h_0$,
$\tilde h_M = Z_{Mh}^{-1} \tilde h_0$,
$t_M = Z_{Mt} t_0$,
$u_M = Z_M u_0$, with
\EA
Z_{Mh} (u_M) & = & 1 - \frac{c}{2 \ep} u_M^2 + O(u_M^4) \hs,
\label{ZMh} \\
Z_{Mt} (u_M) & = & 1 + O(u_M^4) \hs,
\label{ZMt}
\\
Z_M (u_M) & = & 1 + \frac{c}{2 \ep} u_M^2 + O(u_M^4) \hs.
\label{ZM}
\EE
This reparametrization  respects (\ref{ward}) and renders both the response
function
and the growth rate regular as $\ep \to 0$,
\EQ
L^d G_M (0, t_M \! = \! L^2, u_M) =
L^d G_0 (0, t_0 \! = \! L^2) + O (u_M^2 \ep^0, u_M^4) \hs,
\EN
\EQ
L^{2 - \chi_0} \langle \partial_{t_M} h_M \rangle (L, u_M) =
- b \, u_M + O(u_M^3 \ep^0, u_M^5) \hs.
\label{hreg}
\EN

>From (\ref{ZMh}) and (\ref{beta.pert}), one obtains the beta function
\EQ
\beta_M (u_M) = \ep u_M + c u_M^3 + O(u_M^5) \hs.
\label{beta1L}
\EN
For $d > 2$, it has a pair of real-valued unstable fixed points
\EQ
u_M^{\star \;2} = - \frac{\ep}{c} + O(\ep^2)
\label{ustar}
\EN
that describe the roughening transition from the weak coupling to the strong
coupling phase. Relations (\ref{z}) and (\ref{zeta}) then give the critical
exponents \cite{NattermannTang.weakcoupling}
\EQ
z^\star = 2 + O(\ep^2) \hs, \HS \chi^\star = 0 + O(\ep^2)
\label{exp1L}
\EN
satisfying (\ref{scalingrelation}). While higher-order calculations are
cumbersome in the dynamic framework, we will see in the next section that
these values are exact to all orders in perturbation theory.

\resection{The Roughening Transition: Replica \newline Renormalization}

It is well known that
the KPZ equation can be mapped onto a system of {\em directed polymers}
given by the partition function
\EQ
Z = \int {\cal D} r
\exp \left [ - \frac{1}{2 \nu} \int \D t
             \left ( \frac{1}{2} \left ( \frac{\D r}{\D t} \right )^2 -
                     \lambda \eta
             \right )
     \right ] \hs.
\EN
Here $r(t)$ denotes the polymer displacement field in $d$ transversal
dimensions
as a function of the longitudinal ``timelike'' coordinate $t$. The polymer is
subject
to the {\em quenched} random potential $\lambda \eta (r,t)$ that has the
statistics (\ref{eta}) and describes point impurities with short-ranged
correlations.

The height field of Eq.~(\ref{KPZ}) is related to the restricted
partition sum $Z(r,t)$ of all paths ending at a given point $(r,t)$
by the Hopf-Cole transformation
\EQ
h (r,t) = \frac{2 \nu}{\lambda} \log Z(r,t) \hs.
\EN
Hence the {\em disorder-averaged\/} free energy per unit longitudinal length
\EQ
\overline f(L) \equiv - 2 \nu \lim_{T \to \infty} \partial_T \overline{ \log Z}
(T, L)
\EN
in a system of size $T \times L$ (with periodic boundary conditions in
transversal
direction) is proportional to the stationary growth velocity,
\EQ
\overline f = - \lambda \langle \partial_t h \rangle \hs.
\label{growthrate}
\EN

A convenient way to set up perturbation theory is the replicated partition
function
\EQ
Z_p = \int \prod_{\alpha = 1}^p {\cal D} r_\alpha
      \exp \left [ - \frac{1}{4} \int \D t_0
                  \left ( \sum_\alpha
                          \left ( \frac{\D r_\alpha}{\D t_0} \right )^2
                          - \lambda_0^2 \sum_{\alpha < \beta}
                            \delta (r_\alpha - r_\beta)
                  \right )
           \right ] \hs,
\EN
where the the parameters $\nu$ and $\sigma^2$ are again absorbed into
the definition of the canonical variables $t_0$ and $h_0$. Since we are
interested in arbitrary particle numbers $p$, we rewrite the partition
function in second quantization,
\EQ
Z = \int {\cal D} \phi {\cal D} \bar \phi
      \exp \left [ - \frac{1}{4} \int \D t_0 \D^d r
                   \left (
                   \bar \phi (\partial_{t_0} - \partial^2_r ) \phi -
                   \lambda_0^2 \bar \phi^2 \phi^2
                   \right )
           \right ] \hs,
\EN
where $\phi(r, t_0)$ is a complex field. The normal-ordered interaction term
$ - \lambda_0^2 \bar \phi^2 \phi^2 $
is an {\em attractive} pair contact potential. More generally, we define the
normal-ordered $m$-line contact fields
\EQ
\Phi_m (t_0) \equiv \int \D^d r [\bar \phi (r,t)]^m [\phi (r,t)]^m \hs.
\EN

The perturbation series for the free energy
\EQ
f_p (L, u_0^2) = - L^{-2} \sum_{N=1}^\infty \frac{u_0^{2 N}}{4^N N!}
  L^{-2 N \ep} \int \D t_2 \dots \D t_N
  \langle \Phi_2 (0) \Phi_2 (t_2) \dots \Phi_2 (t_N) \rangle_p
\label{fN}
\EN
is a sum involving connected pair field correlations in the $p$-line
sector of the unperturbed theory ($u_0 = 0$). The integrals in
Eq.~(\ref{fN}) are infrared-regularized by the system size $L$; their
ultraviolet singularities are determined by the short-distance
structure of the pair field correlations and have to be absorbed into
the coupling constant renormalization.  Hence consider the asymptotic
scaling of the $N$-point  function $ \langle \Phi_2 (t_{01}) \dots
\Phi_2 (t_{0N}) \rangle $ as the points $t_{01}, \dots,
t_{0N}$ approach each other,
\EQ
t_{0j} - t_{0k} = t \, \tau_{jk} \HS \mbox{and} \HS t / L^2 \to 0
\EN
with $\tau_{jk}$ and the ``center of mass''
$t'_0 = N^{-1} \sum_{j = 1}^N t_{0j}$ remaining fixed.
This is given by the $p$-independent short-distance algebra
\EQ
\Phi_2 (t_{01}) \dots \Phi_2 (t_{0N}) =
\sum_{m = 2}^{N + 1} t^{- (N - m + 1) d / 2}
   \left [ C_N^m (\tau_1, \dots, \tau_{N-2}) \, \Phi_m (t'_0) + \dots
   \right ] \hs,
\label{OPA}
\EN
where $C_N^m$ are scaling functions of the $N - 2$ linearly independent
distance ratios $\tau_{jk}$ and the dots denote terms that are subleading by
positive integer powers of $t / L^2$.
The integration over the relative distances then yields
\EA
\lefteqn{\BS
\int \D t \, t^{N-2} \prod_{l = 1}^{N - 2} \D \tau_l \;
\langle \Phi_2 (t_{01}) \dots \Phi_2 (t_{0N}) \rangle = } \nonumber
\\ & &
\sum_{m = 2}^{N + 1}
\int J_N^m \, t^{m - 3 + \ep (N - m + 1)} \D t \;
\langle \Phi_m (t'_0) \rangle + \dots
\label{IN}
\EE
with
\EQ
J_N^m = \int \prod_{l = 1}^{N - 2} \D \tau_l \, C_N^m (\tau_1, \dots,
\tau_{N-2}) \hs.
\label{J}
\EN

Hence we obtain to one-loop order
\EQ
L^2 f_p (L, u_0^2) = - L^d \langle \Phi_2 \rangle_p \; \cdot \frac{1}{4}
                     u_0^2 \left (1 + \frac{c}{\ep} u_0^2 \right )
+ 0 (u_0^4 \ep^0, u_0^6)
\label{f1L}
\EN
with
$ L^d \langle \Phi_2 \rangle_p = p (p - 1) / 2 $. The pole in (\ref{f1L})
originates from the term in (\ref{IN}) with $m = N = 2$, which is shown
diagrammatically in Fig.~2(a). This pole
can be absorbed into the definition $u_M^2 = Z^2_M u_0^2$ with $Z_M$
given by (\ref{ZM}). To this order, we hence recover the beta function
(\ref{beta1L}) and the fixed points (\ref{ustar}) of the dynamic calculation.

In the polymer framework, however, it is not difficult to discuss higher
orders.
The ultraviolet singularities of the $N$-th order integral (\ref{IN}) are
contained
in the coefficients $J_N^m$ or arise from the integration over $t$. In
the first case, they are due to a {\em proper subdiagram} and hence
already absorbed into the renormalized coupling constant at lower order.
Only the divergences from the integration over $t$, with $J_N^m$
denoting the regular part of (\ref{J}), may contribute to the {\em
primitive singularity} at order $N$. Inspection of (\ref{IN}) then
shows that a pole at $\ep = 0$ only appears for $m = 2$. However, there
is only one diagram per order of this kind, which is shown in Fig.~2(b).
Since this diagram factorizes into loops of the kind of Fig.~2(a), it
contributes a pole in $\ep$ of order $N - 1$. Therefore the pole at
order $N = 2$ is the only primitive singularity in the
series (\ref{fN}) for the free energy; analogous arguments apply to the
expansions of the contact field correlation functions. It follows
that the one-loop equations (\ref{ZM}), (\ref{beta1L}) and
(\ref{ustar}) are exact to all orders in perturbation theory (see also
\cite{BhRa,FERMIONS}).

The replica trick is unproblematic within perturbation theory, since it
reduces to convenient bookkeeping of the averaging over disorder.
Indeed, the random limit $p \to 0$ is trivial in Eqns.~(\ref{ZM}),
(\ref{beta1L}) and (\ref{ustar}) which are  independent of $p$.
The crossover scaling function of the disorder-averaged free energy
\EQ
{\cal C} (u_M^2) \equiv L^2 \overline f (L, u_M^2) =
                        L^2 \lim_{p \to 0} \frac{1}{p} f_p (L, u_M^2)
\EN
is a regular function of the minimally subtracted coupling constant
$u_M^2$, as follows by inserting (\ref{ZM}) into (\ref{f1L}).
By (\ref{growthrate}), (\ref{ZMh}), (\ref{ZMt}), and (\ref{ZM}), the
function $ - {\cal C} (u_M^2) / u_M $ equals the scaling function
(\ref{hreg}) of the minimally subtracted growth rate, and hence $b =
1/8$. If the
renormalization point condition (\ref{hL}) is chosen, the universal
function ${\cal C} (u_M^2)$ is directly related to the renormalized
coupling constant,
${\cal C} (u_M^2) = b \, u_R^2 (u_M^2)$.
Its finite fixed point value
\EQ
{\cal C}^\star = {\cal C} (u_M^{\star \; 2}) =
 - \frac{b}{c} \, \ep  + O(\ep^2) \hs,
\label{Cstar}
\EN
the {\em Casimir amplitude} at the roughening transition, is the analogon
of the central charge in conformally invariant field theories
\cite{BloteAl.c/Affleck.c}.
The function ${\cal C}(u_M^2)$ has the finite limit (\ref{Cstar}) since
the free energy does not develop an anomalous dimension at the roughening
transition, i.e.~hyperscaling is preserved. By
(\ref{growthrate}), this implies that
$\langle \partial_t h \rangle \sim L^{-2}$ at the transition, and hence
$ \chi^\star - z^\star = 2$. The exponents (\ref{chistar.zstar}) then
follow from this relation together with (\ref{scalingrelation}).

The dynamic exponent $z^\star = 2$ can also be verified independently.
Consider the two-point function of the normal-ordered density field
$\Phi_1 (t_0) \equiv \bar \phi \phi (r \! = \! 0, t_0)$,
\EA \lefteqn{
\langle \Phi_1 (t_0) \Phi_1 (t'_0) \rangle (L, u_0^2) = } \nonumber
\\ & &
\sum_{N = 0}^\infty \frac{u_0^{2N}}{4^N N!} L^{-2 N \ep}
  \int \D t_{01} \dots \D t_{0N}
  \langle \Phi_1 (t_0) \Phi_1 (t'_0) \Phi_2 (t_{01}) \dots \Phi_2 (t_{0N})
\rangle \hs.
\label{Rpert}
\EE
Its short-distance asymptotics gives the {\em return probability} of a
single line to the origin $ r = 0 $. In the linear theory,
\EQ
\langle \Phi_1 (t_0) \Phi_1 (t'_0) \rangle (L, 0)
\sim |t_0 - t'_0|^{- d / z_0}
\EN
for $ |t_0 - t'_0| / L^2 \ll 1$. Any anomalous contribution to this exponent
arises from the renormalization of the fields $ \Phi_1 (t_0) $ and $ \Phi_1
(t'_0) $.
The renormalization of $\Phi_1 (t_0)$ is due to a short-distance
coupling of the form
\EQ
\Phi_1 (t_0) \Phi_2 (t_{01}) \dots \Phi_2 (t_{0N}) =
  t^{- N d / 2} C_N^1 (\tau_1, \dots, \tau_{N - 1}) \, \Phi_1 (t_0) + \dots
\hs
\EN
for $t / L^2 \to 0$ (with
$t_{0j} - t_0 = t \tau_{0j}$, $t_{0j} - t_{0k} = t \tau_{jk} $
for $ j,k  = 1, \dots, N $, and  $\tau_1, \dots, \tau_{N-1}$
denoting a basis of the fixed ratios $\tau_{0j}, \tau_{jk}$),
and there is a corresponding expression for $\Phi_1 (t'_0)$.
However, it is obvious that the product on the l.h.s.~couples only to
contact fields of at least two lines, and therefore
$C_N^1 = 0$ at all orders $N$.

\resection{The Strong-coupling Fixed Point}

As discussed in the previous sections, the crossover from the critical fixed
point to the Gaussian fixed point in $d > 2$ can be parametrized in terms of
the coupling constant $u_M$ of a minimal subtraction scheme. In the framework
of the $\ep$-expansion, $u_M$ is completely equivalent to the coupling constant
$u_R$ defined by normalization conditions, e.g.~(\ref{CL}) and (\ref{GL}):
the two couplings are related by  a
diffeomorphism that remains regular in the limit $\ep \to 0$, defined on a
domain of interaction space that contains the fixed points $u_M = u_R = 0$  and
$u_R^\ast (u_M^\ast)$. This equivalence is lost in the crossover to the
strong-coupling fixed  point, as I will now show.

Integration of the flow equation (\ref{beta}) for $u_M^2$ in two dimensions
with
the initial condition $ u_M^2 (L_0) = u_1^2$ yields
\EQ
u_M^2 (L) = \frac{u_1^2}{1 - u_1^2 \log (L/L_0)}  \hs,
\label{g_M}
\EN
an expression that diverges at a finite value $L = 1 / u_1^2$.  It follows
immediately that $u_M (L)$ is not well suited to describe the  crossover to the
strong-coupling fixed point since it divides the crossover  into an ultraviolet
regime $ L < 1 / u_1^2 $ and an infrared regime  $ L > 1 / u_1^2 $ where $u_M^2
(L)$
is defined by analytic continuation of the solution (\ref{g_M}). However, the
pole of $u_M^2 (L)$ at $ L = 1/u_1^2 $ is only a ``coordinate  singularity''
\cite{GEOMETRY} of the minimal subtraction scheme; any renormalized coupling
$u_R (L) $ is expected to remain regular at $L = 1 / u_1^2$. Hence the function
$ u_M^2 (u_R^2) $ also has a pole at the value $ u_R^2 (L \!=\! 1/u_1^2) $
between the
fixed points $u_R^2 = 0$ and $u_R^{\ast \; 2}$.

For small $u_M^2 > 0$, the correlation functions (\ref{CL}) and (\ref{GL})
at the normalization point
can be calculated perturbatively, which results in an analytic renormalization
\EA
h_R  & = & [1 + a_1 u_R^2 + a_2 u_R^4 + \dots] h_M  \hs,
\label{hRpert}
\\
t_R  & = & [1 + b_1 u_R^2 + b_2 u_R^4 + \dots] t_M  \hs,
\label{tRpert}
\EE
and hence by (\ref{scalingrelation})
\EQ
u_M^2  =  [1 + (a_1 + b_1) u_R^2 + (a_2 + a_1 b_1 + b_2) u_R^4 + \dots] u_R^2
\hs.
\label{uMR}
\EN
with finite coefficients $a_N, b_N$. Hence $u_R^2$ has the beta function
\EQ
\beta_R (u_R^2) =
u_R^4 + (a_1^2 + a_1 b_1 + b_1^2 - a_2 - b_2) u_R^8 + O(u_R^{10}) \hs.
\label{betaR}
\EN
In perturbation theory, one would calculate these power series up to some
finite order in $u_R^2$ and look for a fixed point $u_R^{\ast \; 2}$ of
Eq.~(\ref{betaR}). However, a low-order calculation does not yield nontrivial
exponents at the strong-coupling fixed point. Consider
e.g.~the expression (\ref{zeta}) for the roughness exponent in $d = 2$,
\EQ
\chi (u_R^{\ast 2}) =
- \beta_R (u_R^2)
\left. \frac{\D}{\D u_R^2} \log Z_h (u_R^2) \right |_{ u_R^2 = u_R^{\ast \; 2}}
\hs.
\label{zeta2}
\EN
It is zero if $Z_h (u_R^2)$ is a regular function of $u_R^2$ since
$\beta_R (u_R^{\ast \; 2}) = 0$. In an ordinary $\ep$-expansion, the beta
function
and the $Z$-factor are treated as power series; a finite result
would then arise from the linear part of $\beta_R (u_R^2)$ together with
the simple poles of $Z_h (u_R^2)$. Here, if we assume the existence of an
expansion
parameter $\tilde \ep$ and naively take $u_R^\ast \sim \tilde \ep$, we obtain
$\chi (u_R^{\ast \; 2}) = O(\tilde \ep^{10})$ from (\ref{hRpert}),
(\ref{betaR}), and (\ref{zeta2}).
In fact, we expect a perturbative fixed point of Eq.~(\ref{betaR}) to be
spurious
at any order, i.e.~not to reflect properties of the large-distance asymptotic
regime. The reason is that all coefficients $a_N$, $b_N$ are finite for $\ep =
0$ and thus depend on details of the infrared regularization, unlike the
residues of the poles in $\ep$ which determine the $Z$-factors in an ordinary
$\ep$-expansion. It is difficult to see how this dependence could cancel out
to produce universal exponents.

In the infrared regime $L > L_0$, the relationship between $u_M^2$ and $u_R^2$
is no
longer given by Eq.~(\ref{uMR}). As $L \to \infty$, the asymptotic behavior of
the
renormalized quantities is
\EQ
u_R^{\ast \; 2} - u_R^2 \sim L^{-y'} = \exp [y' / u_M^2] \hs,
\EN
\EQ
t_R = Z_t t_M \sim L^{z - z_0} = \exp [-(z - z_0) / u_M^2] \hs,
\EN
\EQ
h_R = (u_M^2 u_R^{-2} Z_t^{-1}) h_M \sim ( u_M^2 \exp [(z - z_0) / u_M^2] h_M
\hs,
\EN
in terms of their minimal subtraction counterparts (the exponent $y'$ is
defined by
$ \beta_R (u_R^2) = - y' (u_R^2 - u_R^{\ast \; 2}) + O(u_R^2 - u_R^{\ast
\;2})^2 $).
Hence all of these quantities have
essential singularities at $u_M = 0$, which are tied to an essential
singularity
\EQ
\langle h_M (r_1, t) h_M (r_2, t) \rangle \sim
 u_M^{-4} \exp [-2 (z - z_0) / u_M^2]
\EN
of the correlation function in minimal subtraction in the same limit, for fixed
distance $|r_1 - r_2|$.
This shows the inequivalence of the two renormalization schemes.

\resection{Discussion}

In this paper, it has been shown that the perturbation theory for the KPZ
roughening transition can be developed consistently in the dynamic and in the
directed polymer framework. To all orders, it predicts the exponents
\EQ
z^\star = 2 \hs, \hspace{1cm} \zeta^\star = 0
\EN
resulting from the exact beta function (\ref{beta}) in minimal subtraction.

This perturbation theory gives some indications on the existence of an
upper critical dimension $d_>$ for the KPZ equation. At the critical
fixed point $u_M^{\star \;2}$, the nonlinearity is a relevant perturbation
\EQ
\beta (u_M^2) = - 2 \ep (u_M^2 - u_M^{\star \; 2}) + \dots \hs,
\EN
and the interface roughness changes to leading order in perturbation
theory,
\EQ
\chi (u_M^2) =  (u_M^2 - u_M^{\star \; 2}) + \dots \hs.
\EN
With the assumption that $\chi (L)$ is a monotonic function of $L$, this
implies $\chi > 0$ in the strong-coupling phase. This argument, however,
can only be trusted for $d < 4$. The minimally subtracted pair contact field
$\Phi_{2,M}$ has the exact dimension $2 + 2 \ep$ at the
critical fixed point. This dimension would turn negative for $d > 4$, which
is obviously impossible. The regularization of the perturbation series
(\ref{fN})
by minimal subtraction breaks down at $d = 4$ since already the two-loop
integral $\int \D t_{02} \langle \Phi_2 (t_{01}) \Phi_2 (t_{02}) \rangle$
develops a new pole that is not absorbed into the coupling constant
renormalization
(\ref{ZM}). This exhibits the singular r\^ole of dimension $4$ and suggests the
identification $d_> = 4$, in agreement with recent mode-coupling results
\cite{MooreAl}. It should be possible to corroborate these results by extending
the analysis of the roughening transition to dimensions $d > 4$.

The form (\ref{beta}) of the flow equation has an interesting
consequence for the desription of the crossover to strong-coupling
behavior in $d = 2$. The minimal subtraction coupling constant $u_M^2
(u_R^2)$ develops a pole at a finite value $u_R^2 = u_1^2  < u_R^{\ast \; 2}$
of
the renormalized coupling constant, a singularity not taken into
account by existing renormalization group treatments of the KPZ
equation. That singularity divides the crossover into an ultraviolet
regime $u_R^2 < u_1^2$ and an infrared regime $u_R^2 > u_1^2$. Only the former
is accessible in perturbation theory: even the summation of an infinite
number of terms e.g. in the series (\ref{hRpert}) and (\ref{tRpert})
can only lead to a relation $u_R^2 (u_M^2)$ for $u_M^2 > 0$. Whether
information on the infrared regime can be gained by inverting this
relation and then analytically continuing it to values $u_R^2 > u_1^2$
remains to be seen.

The precise form (\ref{beta}) of the beta function is clearly due to the fact
that the intrinsic dimension of the polymer, or the dimension of time, is one.
The disorder problem, however, has a natural generalization from directed lines
to $D$-dimensional directed surfaces. For $D \neq 1$, higher-order perturbative
singularities are expected, which generate further terms in (\ref{beta}). This
generalization has an interest in its own right. It is not clear, however, if
it is useful to solve the original problem: any perturbative fixed point
will tend to infinity as $D \to 1$, but the exponents may still be well-behaved
in that limit.

Useful discussions with H.~Kinzelbach are gratefully acknowledged.

\newpage

\noindent {\bf Figure Captions}

\begin{enumerate}

\item
Diagrammatic expansions generated by the dynamic functional
(\ref{dynaction}).
The lines with one and two arrows denote the unperturbed  response function
(\ref{G0}) and the unperturbed correlation function (\ref{C0'}), respectively.
Dots represent the vertices $i \tilde h (\nabla h)^2$;
each incoming line to a vertex has to be differentiated
with respect to $r$.
(a) Response function $G_0 (r, t_0, u_0)$. The one-loop diagram is
regular at $\ep = 0$.
(b) Stationary growth rate $\langle \partial_{t_0} h_0 \rangle (L, u_0)$.
The boxed subdiagram contains a simple pole at $\ep = 0$.

\item
Singular diagrams contributing to the finite-size free energy $f_p (L, u_0^2)$
in the
expansion (\ref{fN}).
The lines denote unperturbed single-line propagators \linebreak
$\langle \bar \phi (r_1, t_{01}) \phi (r_2, t_{02}) \rangle$,
the dots pair contact vertices $\Phi_2 (t_0)$.
(a) One-loop diagram containing a primitive pole at $\ep = 0$.
(b) $N$-loop diagram containing a pole of order $N - 1$ at $\ep = 0$.

\end{enumerate}

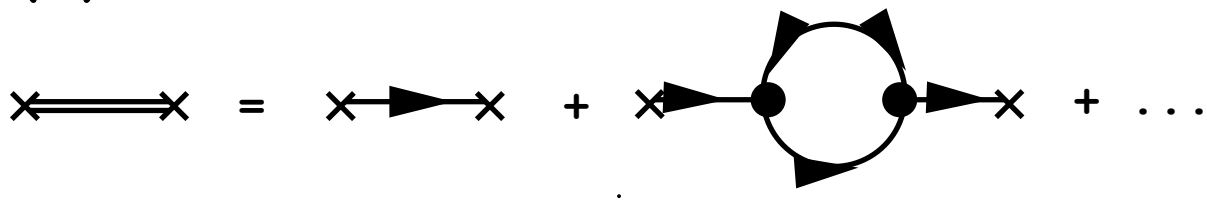
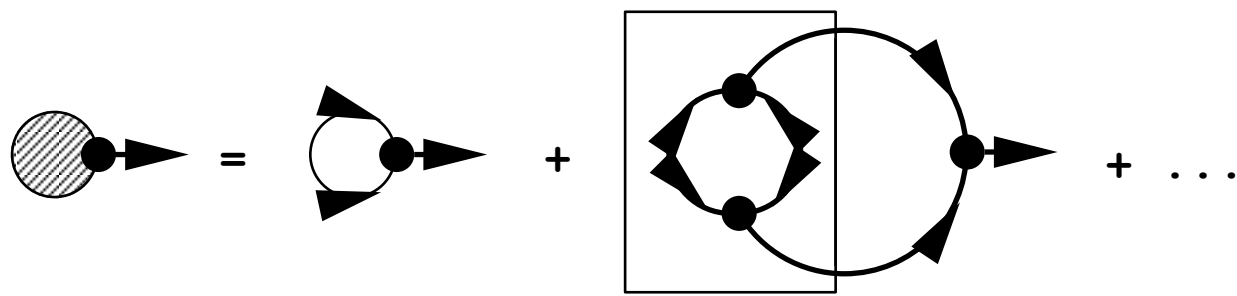
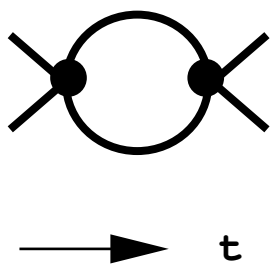
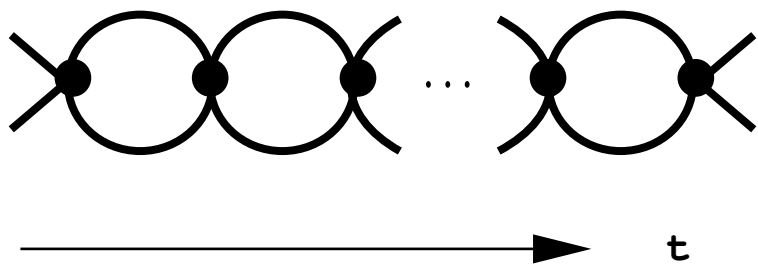
\end{document}